\documentclass[a4paper,11pt]{article}
\usepackage{pos}

\title{Double Deeply Virtual Compton Scattering at Jefferson Lab Hall A}

\author*[a]{Marie Bo\"er}
\author[a]{Debaditya Biswas}

\affiliation[a]{Virginia Tech,\\
  Blacksburg, VA, USA. }

\emailAdd{mboer@jlab.org}

\abstract{
This paper presents our project and perspectives to measure for the first time 
beam spin asymmetries from Double Deeply Virtual Compton Scattering 
in the $eP\to e'P' \mu^+\mu^-$ reaction at Jefferson Lab. 
Our goal is to constrain the so-called Generalized Parton Distribution (GPDs) in a kinematic region that isn't accessible 
from other reactions, such as Deeply Virtual Compton Scattering, to allow for their extrapolation to "zero skewness", i.e. at 
a specific kinematic point enabling for tomographic interpretations of the nucleon's partonic structure.
We are discussing DDVCS phenomenology and our approach, as well as our experimental project aimed at complementing the SoLID experiment at JLab Hall A with 
a new muon detector.
}

\FullConference{25th International Spin Physics Symposium (SPIN 2023)\\
 24-29 September 2023\\
 Durham, NC, USA\\}

\begin{document}
\maketitle

\section{Introduction}

The so-called Generalized Parton Distribution (GPDs)~\cite{Mul94, Ji:1996nm}
are matrix elements parametrizing the soft structure of the nucleon in “Hard Exclusive” reactions~\cite{Ji97,Rad97}. 
GPDs contain information on the longitudinal momentum versus transverse position of the partons (quarks and gluons)~\cite{Die03,Bel05}. 
We have been studying GPDs for the last $\sim$30 years as we are looking to move towards multidimensional images of the nucleon structure. 
One of the interesting interpretations of GPDs is the possibility to access tomographic views of the nucleon, where we can relate the transverse position of the partons to the quark and gluon densities~\cite{Bur02}. 
This kind of interpretation relies on extrapolations of GPDs to certain kinematics that can’t be accessed experimentally, and on models, referred to as “zero skewness”~\cite{Guo23}, 
i.e. reactions where all the momentum transferred to the nucleon is purely transverse. Our goal is to study Double Deeply Virtual Compton Scattering to constrain the GPDs at this limit. \smallskip

\begin{figure}
    \centering
    \includegraphics[width=0.55\linewidth]{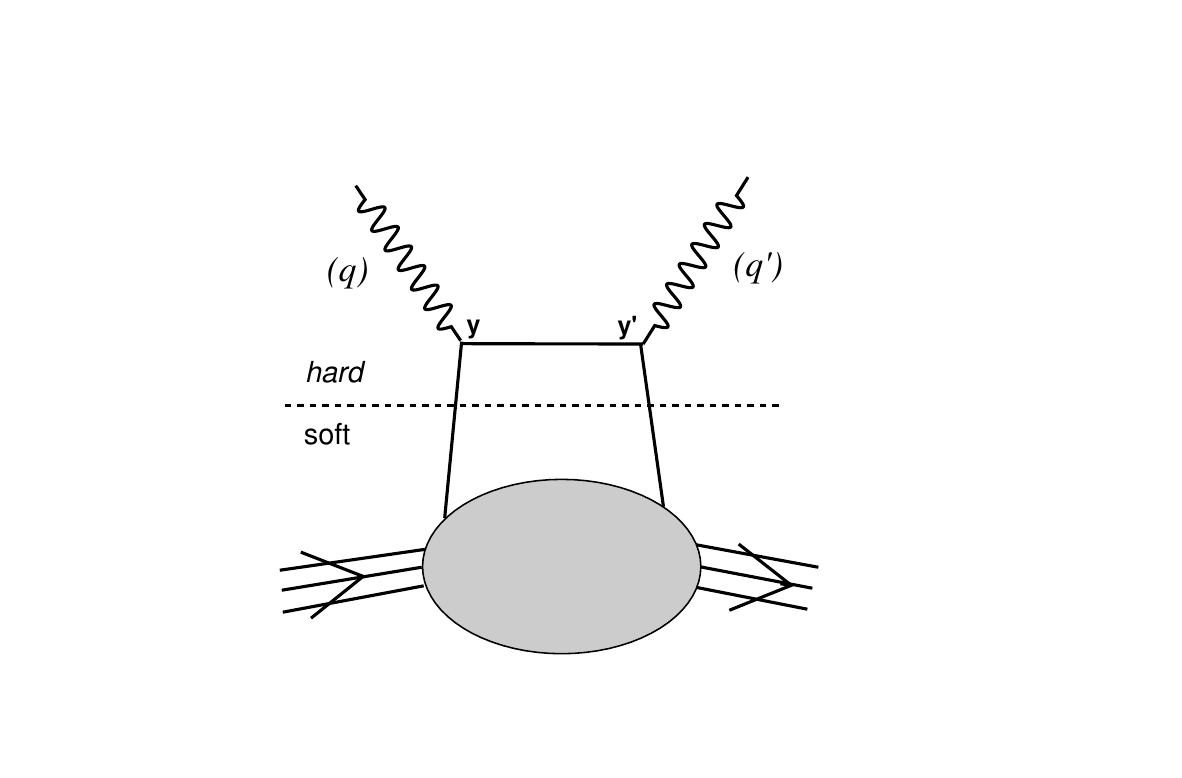} \vspace{-0.9cm}
    \caption{\small{General Compton-like reaction diagram at leading twist, leading order. Dash lane indicates the factorization 
    between a soft part, parameterized by the GPDs and a hard calculable part.}}
\label{fig:genCompton} 
\end{figure}

“Hard Exclusive” reactions refers to: a “hard scale” of at least 1 GeV$^2$, allowing for factorization between a soft part parametrized by the GPDs, 
and a hard part, calculable~\cite{Collins}; 
“exclusivity” refers to all products of the reaction being known, enabling measurement of the total momentum transfer to the nucleon 
(we use Mandelstam variable “t”, the squared momentum transfer). Fig.\ref{fig:genCompton} is the general Compton-like process, where a photon is scattered off a quark in the nucleon. We display the factorization lane, the bottom part representing the GPDs. 
The incoming and scattered photons have to be of different virtuality to allow for a non-zero momentum exchange to the nucleon. 
We can distinguish between 3 particular cases of “Compton Scattering”: 
Deeply Virtual Compton Scattering (DVCS), where the incoming photon is virtual (spacelike) and the outgoing one is real; 
Timelike Compton Scattering (TCS), where the incoming photon is real and the outgoing one is virtual, subsequently decaying into a lepton pair; 
and Double Deeply Virtual Compton Scattering (DDVCS), where both photons are virtual. 
DVCS has been measured at multiple facilities~\cite{Adl01,Air01,Ste01,Che03,Che06,Mun06,Air07,Air08,Gir08,Air10,Air11,Jo15,Def15,Say18,Akh19,Geo20}, 
TCS has recently been measured for the first time at JLab~\cite{TCSB},  DDVCS has never been measured. \smallskip

There are several GPDs, for quarks and gluons, and to account for relative helicity states of the quark-nucleon system. At leading order and leading twist (lowest order in photon’s virtuality related to extra-gluon exchanges), for a spin 1/2 nucleon, we have 4 (x2 for quarks and gluons) chiral-even GPDs, and 4 (x2) chiral-odd GPDs 
(with quark helicity flip), i.e. 16 total (see for instance ~\cite{Die03,Guidal:2013rya}). 
These GPDs depend on 3 variables: t, x (nucleon’s longitudinal momentum fraction carried by the parton), $\xi$ (“skewness”, related to the longitudinal momentum transfer to the quark in light cone frame). 
We will neglect here their evolution with the photons’ virtuality (namely Q$^2$=-q$^2$ and/or Q’$^2$=q’$^2$ for incoming and outgoing photons, respectively, 
defined from their squared 4-momenta). 
GPDs can’t be measured directly: we measure Compton Form Factors (CFFs), functions of the GPDs, accessible from fits of cross sections and asymmetries of the various reactions. Most models are currently constrained by measurements of DVCS only, where GPDs can only be accessed at specific kinematic points, for x=$\pm\xi$. TCS being the “time-reversal” equivalent of DVCS at leading order and leading twist~\cite{Ber02}, it accesses GPDs at the same kinematics. On the other hand, we can vary the relative virtualities of the two photons in DDVCS to access different kinematics, such as $|x|< \xi$~\cite{Die03,Gui03,Bel03}. It is essential to deconvolute these 2 variables and extrapolate the GPDs to $\xi$=0~\cite{Guo23}, which is needed for tomographic interpretations. \smallskip

\begin{figure}[!ht]
\centering
\includegraphics[width=7 cm]{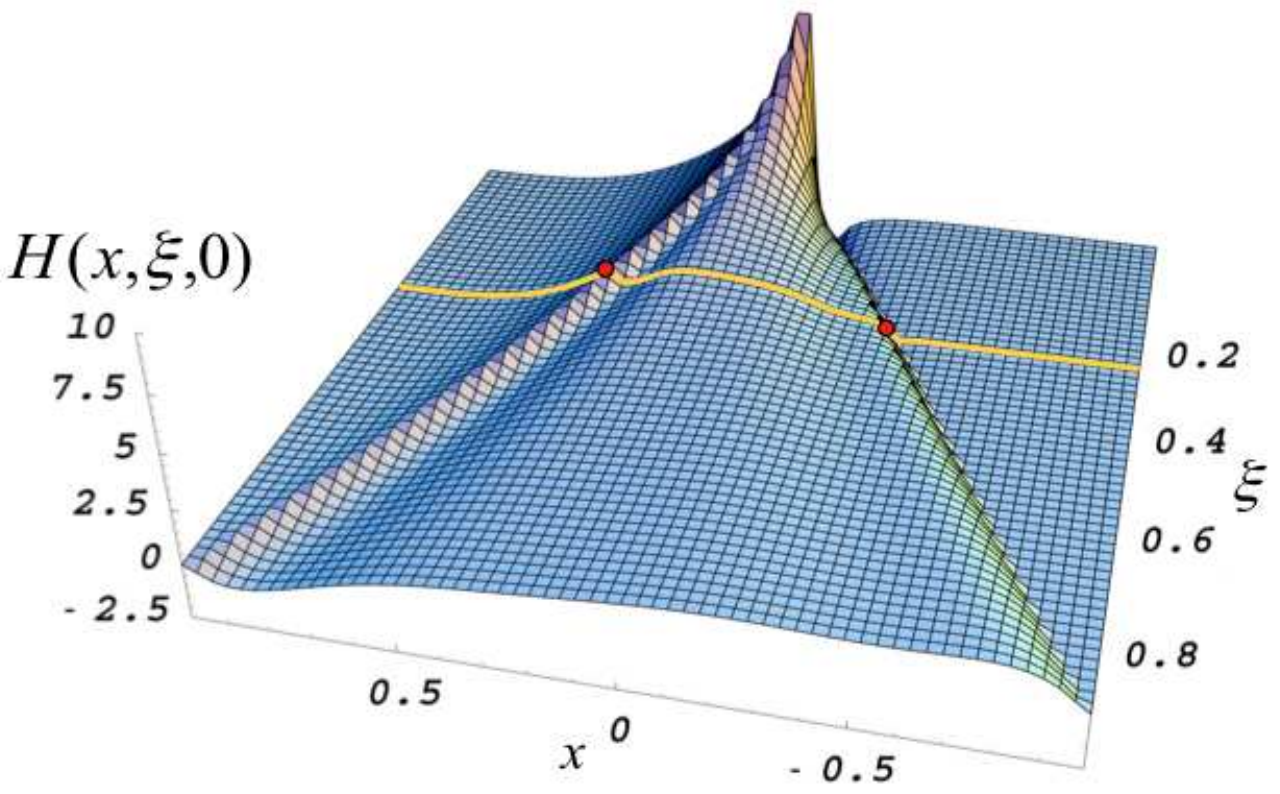}\vspace{-0.1cm}
\caption{Graphical representation of the DVCS Compton form factor (CFF) showing a typical model for the GPD $H$ at $t$=0; the red 
points indicates the GPD values involved in the CFF imaginary part, and the yellow line underlines the integral path of the CFF 
real part.}
\label{fig:CFF}
\end{figure}

Indeed, one can't directly access the GPDs with DVCS, TCS or DDVCS: we are measuring functions of them, the Compton Form Factors (CFFs), which can be extracted from fitting experimental observables (cross sections, 
spin or charge asymmetries...). For instance, the CFF ${\cal H}$ associated with the GPD $H$ and accessible in DVCS or TCS experiments writes 
\begin{equation}
{\cal H}(\xi,t) = \sum_q e_q^2 \left\{ { {\cal P} \int_{-1}^{1} d x \, H^q(x,\xi,t) \, \left[ \frac{1}{\xi - x} - \frac{1}{\xi + x} \right] 
+ i \pi \, { \left[ H^q(\xi,\xi,t) - H^q(-\xi,\xi,t) \right] } } \right\} 
\label{eq:dvcs}
\end{equation}
where the sum runs over all parton flavors with elementary electrical charge $e_q$, and $\cal P$ indicates the Cauchy principal 
value of the integral. 
The imaginary part of the CFF accesses the GDP values at $x=\pm \xi$ and the real part comes in an integral over x. 
It involves the convolution of parton propagators and the GPD values out-of the diagonals $x=\pm \xi$ (Fig.~\ref{fig:CFF}). 
Thanks to the virtuality of the final state photon (see section\ref{sec:pheno}), DDVCS provides a way to circumvent this limitation~\cite{{Gui03},{Bel03}}, 
allowing to vary independently $x$ and $\xi$. Considering the same GPD $H$, the corresponding 
CFF for the DDVCS process writes
\begin{equation}
{\cal H}(\xi',\xi,t) = \sum_q e_q^2 \left\{ { {\cal P} \int_{-1}^{1} d x \, H^q(x,\xi,t) \, \left[ \frac{1}{\xi' - x} - \frac{1}{\xi' 
+ x} \right] + i \pi \, { \left[ H^q(\xi',\xi,t) - H^q(-\xi',\xi,t) \right] } } \right\}
\end{equation}
involving an additional scaling variable $\xi$ representing here the GPD skewness. We can therefore explore the GPDs out of the "diagonal" $x=\pm\xi$ (Fig .\ref{fig:CFF}). 

\section{Phenomenology of DDVCS}
\label{sec:pheno}

\begin{figure}
    \centering
    \includegraphics[width=0.42\linewidth]{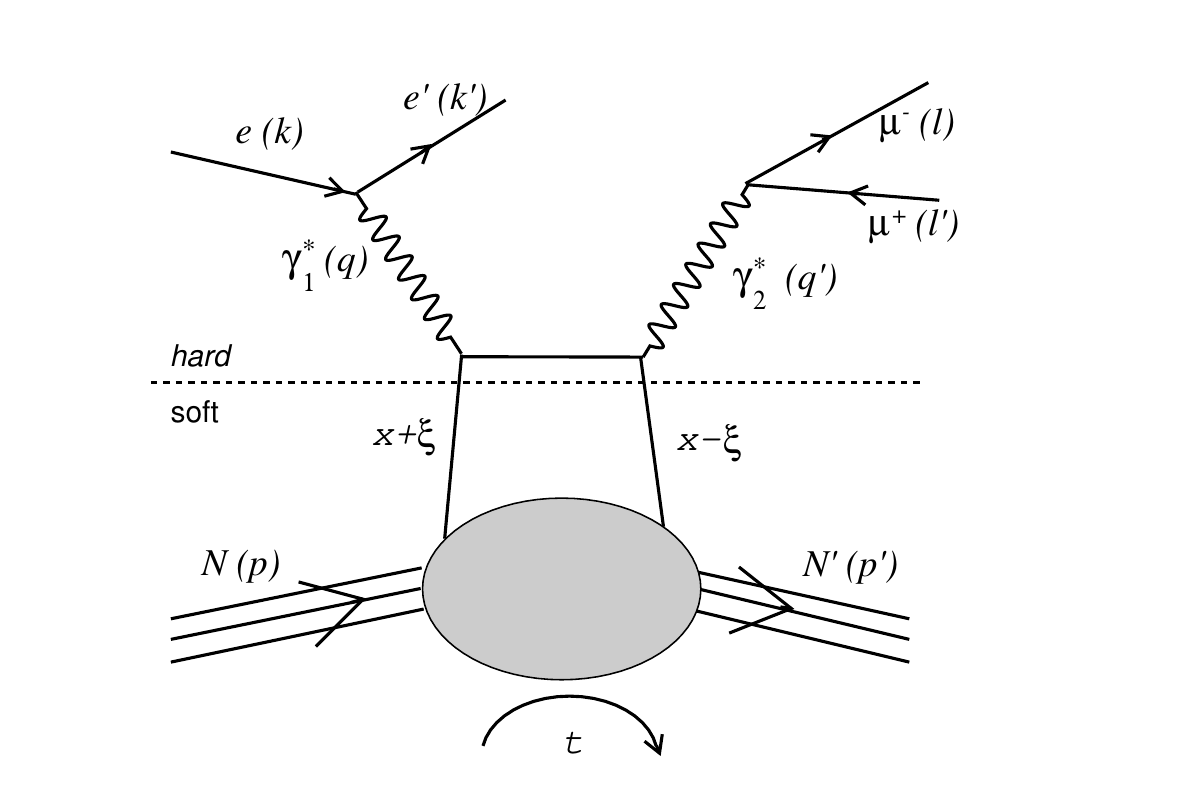}\\
    \includegraphics[width=0.42\linewidth]{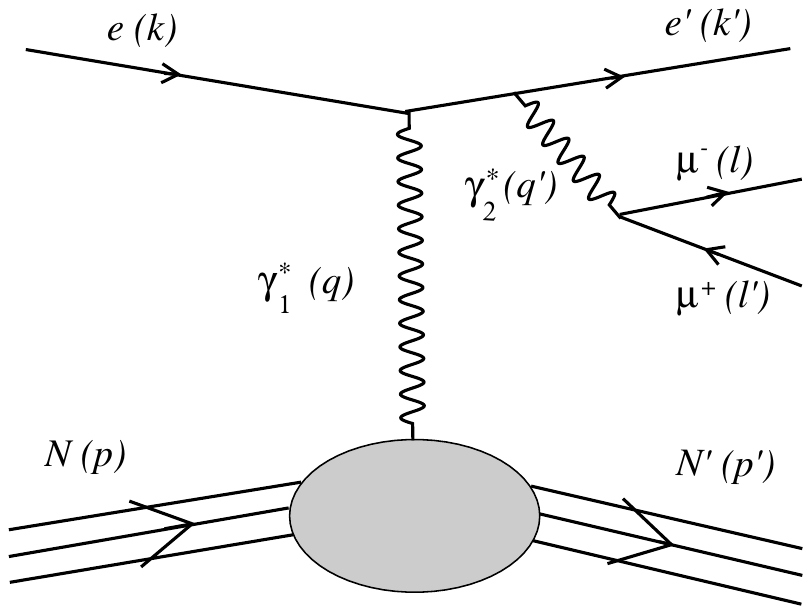}
    \includegraphics[width=0.42\linewidth]{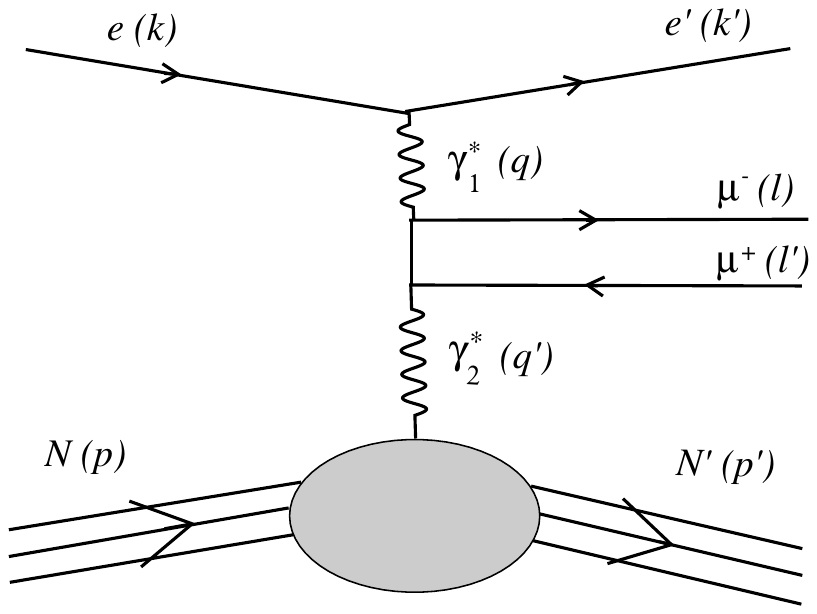} \vspace{-0.5cm}
    \caption{Electroproduction of a muon pair off a quark in a nucleon, at leading order and leading twist (CC not represented). Top: DDVCS process. Bottom left: Bethe-Heitler like process from a radiative virtual photon emission (referred to in this paper as "BH1". Bottom right: Bethe-Heitler process from exchanged virtual photon conversion (referred to here as "BH2"). We indicated some kinematic variables and 4-momenta as 
    indicated in the text.}
    \label{fig:DDVCSBH}
\end{figure}

DDVCS is measured in the reaction $e N\to e' N' L^+L^-$, where N is a nucleon, e is a lepton (beam),  $L^+L^-$ is a lepton pair coming from the decay an outgoing virtual photon. 
At the quark level, the process which occur for DDVCS is: $\gamma_1 N \to \gamma_2 N'$ (as in Fig.\ref{fig:genCompton}). 
In fact, DDVCS interfers with two Bethe-Heitler like processes ("BH" = BH1 and BH2), where the lepton pair comes from (1) a radiative virtual photon emission from the lepton beam ("BH1") and (2) a lepton pair produced by the exchanged photons ("BH2"). 
We are representing the different subprocesses Fig.\ref{fig:DDVCSBH}: the DDVCS process is displayed at the top, we took the case of an electron beam and a final muon pair; we are displaying the two cases of Bethe-Heitler at the bottom ("BH1" on the left, "BH2" on the right). Note that crossed diagrams aren't represented. At leading twist and leading order, 6 diagrams interfer in the 
reaction of electroproduction of a muon pair off a nucleon. We are choosing to study DDCCS+BH in the di-muon channel to avoid concerns in having two electron in the final state (antisymmetrization for the interpretation, particle identification and kinematic reconstruction at the experimental level...), since our available beam at Jefferson Lab (JLab) is an electron beam. \smallskip

\begin{figure}[!ht]
\centering
  \includegraphics[width=0.6\linewidth]{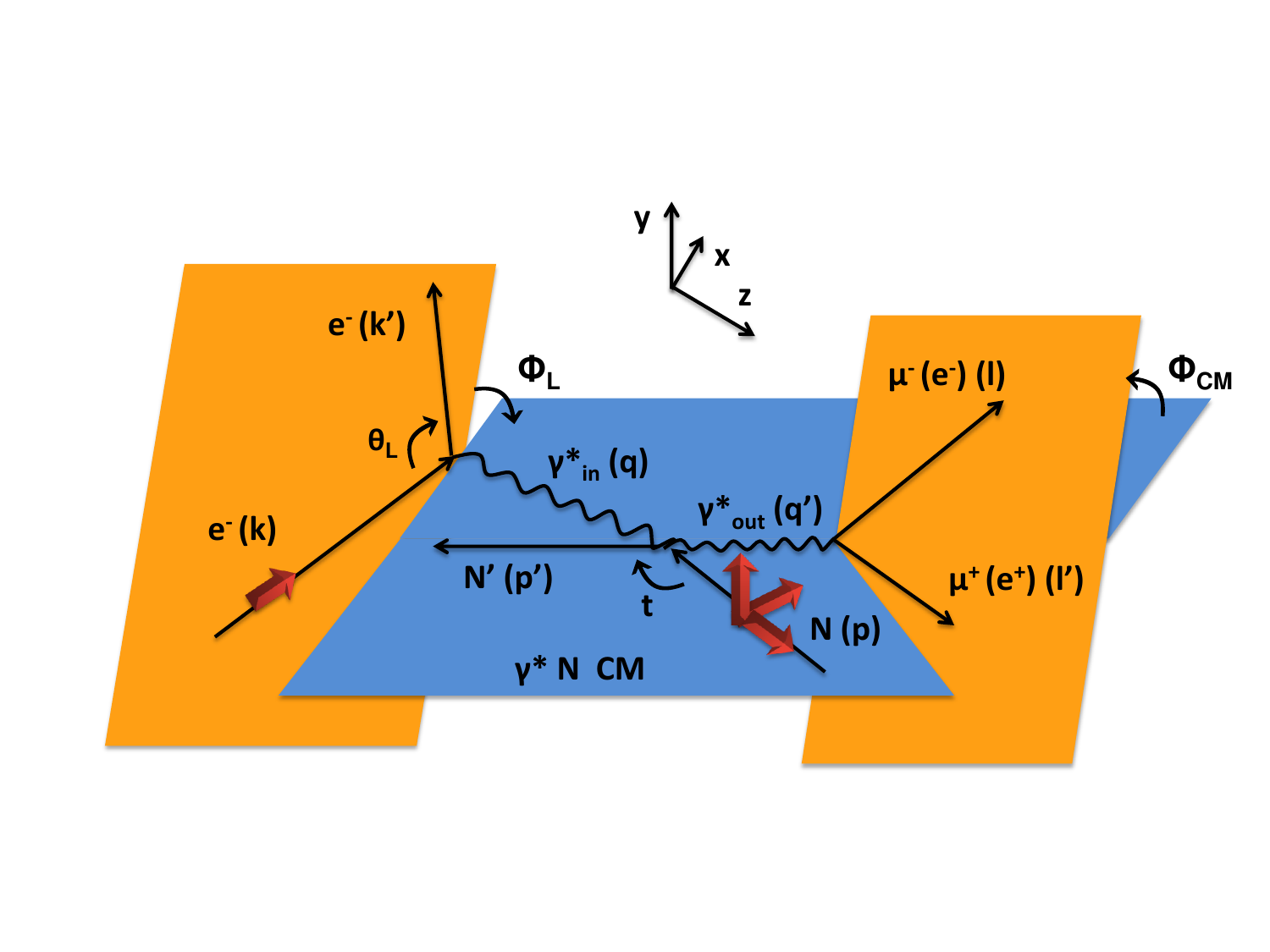}
  \vspace{-1.2cm}
\caption{Reference frame, kinematics, angles, for the DDVCS reaction, using notations defined in~\cite{Gui03} and our angle's notation.} 
\label{fig:frames}
\end{figure}

As indicated in Fig.\ref{fig:DDVCSBH}, 
we are using here the following notation for the 4-vectors of the different particles involved: 
\begin{equation}
e(k) -  e'(k') + p(p_1) \equiv \gamma^{\star}(q_1) + p(p_1) \to p'(p_2) + \gamma^{\star}(q_2) \to p'(p_2) + {\mu^+}(l^+) + 
{\mu^-}(l^-).
\end{equation}
For our kinematics, we are using the notations  defined in~\cite{Gui03}, expressed in the reference frame represented  Fig.~\ref{fig:frames}.
 The photon virtualities (incoming and outgoing, respectively) are defined as
\begin{equation}
Q^2 = - q^2; \hspace{3mm}
Q'^2 = q'^2.
\end{equation}
We define  the symmetrical momentum variables $p$ and $q$ as 
\begin{equation}
q = \frac{1}{2} (q + q');\hspace{3mm}
\hspace{3mm}
p = p + p'   ,
\end{equation}
and the four-momentum transfer to the nucleon $\Delta = p - p' = q - q'$ with $t=\Delta^2$.
 The DDVCS scaling variables  write
\begin{equation}
\label{SymmetricVariables}
x_B = - \frac{1}{2} \, \frac{q_1 \cdot q_1}{p_1 \cdot q_1};\hspace{3mm}
\hspace{3mm}
\xi' = - \frac{q \cdot q}{p \cdot q};\hspace{3mm}
\hspace{3mm}
\xi = \frac{\Delta \cdot q}{p \cdot q}
.
\end{equation}The symmetrical momentum q can be decomposed as 
\begin{equation}
q^2 = - \frac{1}{2} \, {\left( Q^2 - Q'^2 + \frac{\Delta^2}{2} \right) } \, .
\end{equation}
We can decompose the skewness variables in terms of virtualities and $\Delta$:
\begin{equation}
\label{XiEta}
\xi = \frac{Q^2 - Q'^2 + (\Delta^2/2)}{2(Q^2/ x_B) - Q^2 - Q'^2 + \Delta^2}
\, , \qquad
\xi' = - \frac{Q^2 + Q'^2}{2(Q^2/ x_B) - Q^2 - Q'^2 + \Delta^2}
\, ,
\end{equation}
which expresses GPDs variables of interest ($\xi$, $\xi'$, t) in terms of experimentally measured quantities. 
Note that the different $Q'^2$-dependence in the numerators of $\xi$ and $\xi'$ expresses the ability to access out-of diagonals phase space ($x\neq\xi$) from the "lever arm" in Q$^2$ versus Q'$^2$; 
and that for DVCS or TCS, $\xi=\xi'$ (at the asymptotic limit).   \smallskip

\begin{figure}
    \centering
    \includegraphics[width=0.7\linewidth]{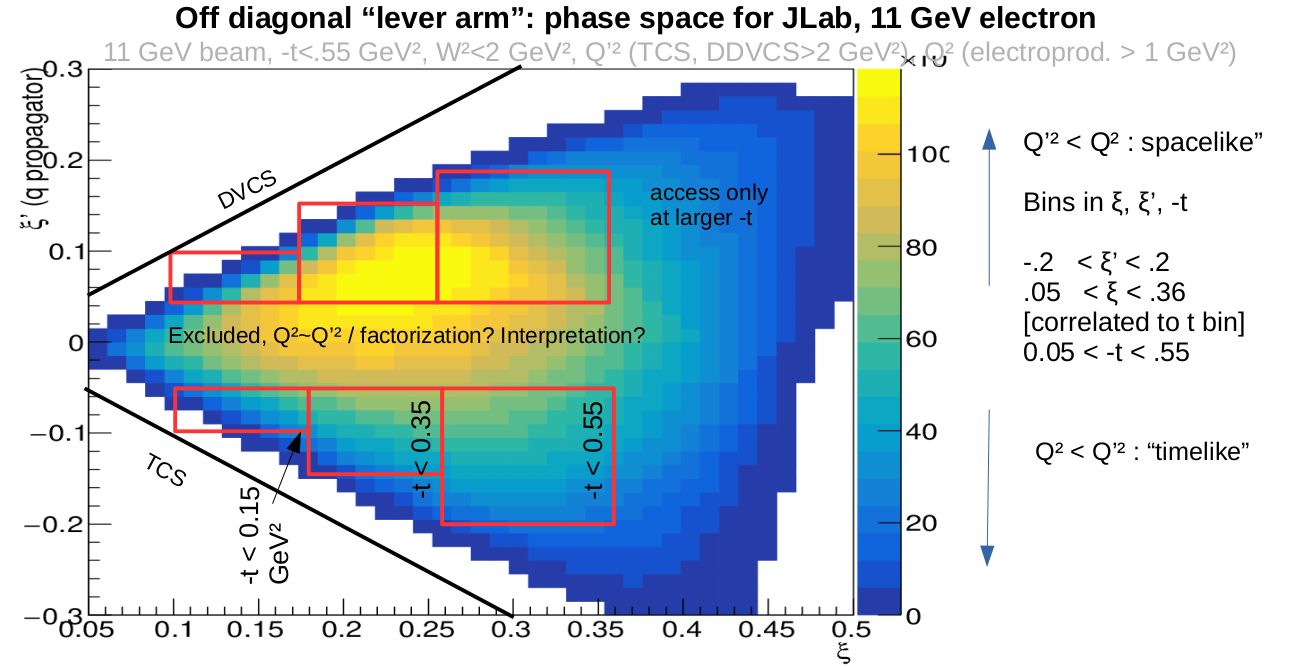}
    \caption{Accessible "off-diagonal" phase-space for DDVCS+BH for experiments at Jefferson Lab. Cuts are indicated on the figure and the main text. Note that the figure (blue/yellow area) integrates our statistics over t:     red boxes represent bins in t, where the maximal limits are indicated in the boxes. }
    \label{fig:psdd}
\end{figure}

We made projections for realistic kinematics accessible at Jefferson Lab, using an 11 GeV electron beam, cuts at $Q^2$=1~GeV$^2$, $Q'^2$=4~GeV$^2$, $Q^2-Q'^2$=1~GeV$^2$, $W^2>2$GeV$^2$, 
$-t<0.55$GeV$^2$, among other kinematic considerations. Neglecting the acceptance effects, we obtained Fig.\ref{fig:psdd}, showing how much "off-diagonal" phase space we would be able to cover 
in an experiment. Red boxes on this figure indicate one proposed binning in t, $\xi$ and $\xi'$. We decided to exclude the 
central region ($Q^2\sim Q'^2$) as it can't be interpreted with our current knowledge (factorization may not hold). 

\section{DDVCS+BH observables and projections}

 We define as "experimental observables" unpolarized, beam and/or target polarized cross sections and spin asymmetries (normalized difference of 2 relative spin states). At JLab, we would like to 
 measure unpolarized and beam polarized cross sections. CFFs can be extracted from fitting these cross sections. 
The differential cross section for the electroproduction of muon pair 
off the nucleon can be written~\cite{Bel03-1} 
\begin{equation}
\frac{d^7 \sigma}{dx_B \, dy \, dt \, d\phi \, dQ'^2 \, d\Omega_{\mu}} = 
\frac{1}{(2\pi)^3} \, \frac{\alpha^4}{16} \, \frac{x_B y}{Q^2 \sqrt{1+\varepsilon^2}} \, 
\sqrt{1 - \frac{4 m_{\mu}^2}{Q'^2}} \, \, { \vert \mathcal{T} \vert }^2 ,
\end{equation}
where the reaction amplitude can generically be expressed as
\begin{equation}
{ \vert \mathcal{T} \vert }^2 = { \vert \mathcal{T}_{VCS} \vert }^2 + \mathcal{I}_1 + \mathcal{I}_2 + { \vert \mathcal{T}_{BH_1} 
\vert }^2 + { \vert \mathcal{T}_{BH_2} \vert }^2 + \mathcal{T}_{BH_{12}} ,
\end{equation}
with the pure DDVCS amplitude ${\vert \mathcal{T}_{VCS} \vert }^2$, the interference amplitudes $\mathcal{I}_1$ and $\mathcal{I}_2$ 
between the DDVCS and two Bethe-Heitler sub-processes, and the pure BH amplitude built itself from the two elementary BH subprocesses. We refer to~\cite{Bel03-1}
 for more details on the harmonic structure of these amplitudes. \smallskip

The interference amplitude bewteen the BH and DDVCS processes is an observable of interest since 
it involves linear combinations of Compton form factors.
The imaginary part of the amplidudes can be accessed  directly via
beam spin asymmetries. Not only beam spin asymmetries are easier to measure than cross sections, 
but "pure" Bethe-Heitler terms in the amplitude cancel in beam spin asymmetries (at the numerator), leading to a better sensitivity to the CFFs and GPDs. 
 Considering the harmonic dependence of the cross section, it was shown~\cite{Bel03-1} that the 
same basic information about GPDs can be obtained from the appropriate moments in $\phi$ or $\varphi_\mu$, a feature of particular 
interest for experimental consistency. Taking advantage of the symmetry properties of the BH propagators to minimize the BH 
contribution, the first $\phi$-moment and $\varphi_\mu$-moment of the beam spin asymmetry can be written~\cite{Bel03-1}
\begin{eqnarray}
\left\{
A_{\rm LU}^{\sin\phi} 
\atop
A_{\rm LU}^{\sin\varphi_{\mu}}
\right\}
&\!\!\!=\!\!\!&
\frac{1}{{\cal N}}
\int_{\pi/4}^{3\pi/4}\! d \theta_{\mu}
\int_{0}^{2\pi}\! d \varphi_\mu
\int_{0}^{2\pi}\! d \phi \, 
\left\{ 2 \sin\phi \atop 2 \sin\varphi_\mu \right\}
\frac{d^7\overrightarrow{\sigma} - d^7\overleftarrow{\sigma}}{dx_B \, dy \, dt \, d\phi \, dQ'^2 \, d\Omega_{\mu}}
\nonumber\\
\!\!\!&\propto&\!\!\!
\Im{\rm m}
\left\{
F_1 {\cal H}
-
\frac{t}{4 M_N^2} F_2 {\cal E}
+
\xi' (F_1 + F_2) \widetilde {\cal H}
\right\} \, , \label{eq:ALU}
\end{eqnarray}
where $F_1$ and $F_2$ are the proton's form factors, $\cal{H,\tilde{H},E}$ are the CFFs depending on GPDs H, $\tilde{H}$, E, respectively, 
and with the normalization factor given by
\begin{equation}
{\cal N} = \int_{\pi/4}^{3\pi/4}\! d \theta_{\mu} \int_{0}^{2\pi}\! d \varphi_\mu \int_{0}^{2\pi} d \phi \, 
\frac{d^7\overrightarrow{\sigma} + d^7\overleftarrow{\sigma}}{dx_B \, dy \, dt \, d\phi \, dQ'^2 \, d\Omega_{\mu}} \, ,
\end{equation}
and where we omit for clarity the $(\xi',\xi,t)$-dependence of the CFFs. In the case of a proton target the measurement gives 
access to the "out-of diagonal" kinematic dependencies of GPD $H$ (dominant term). \smallskip

\begin{figure}[!ht]
\begin{center}
\includegraphics [width=0.45\textwidth,angle=0]{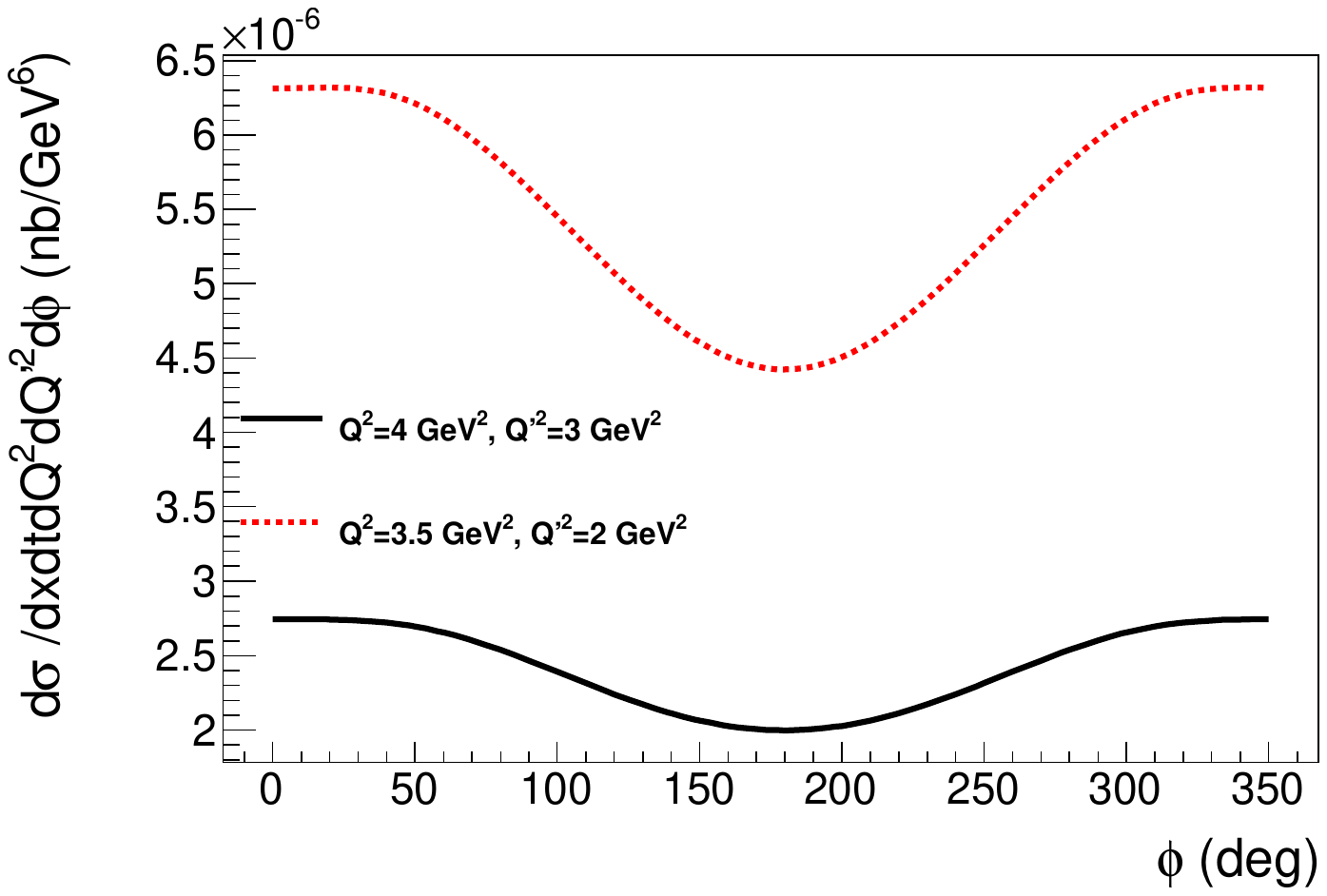} 
\includegraphics [width=0.45\textwidth,angle=0]{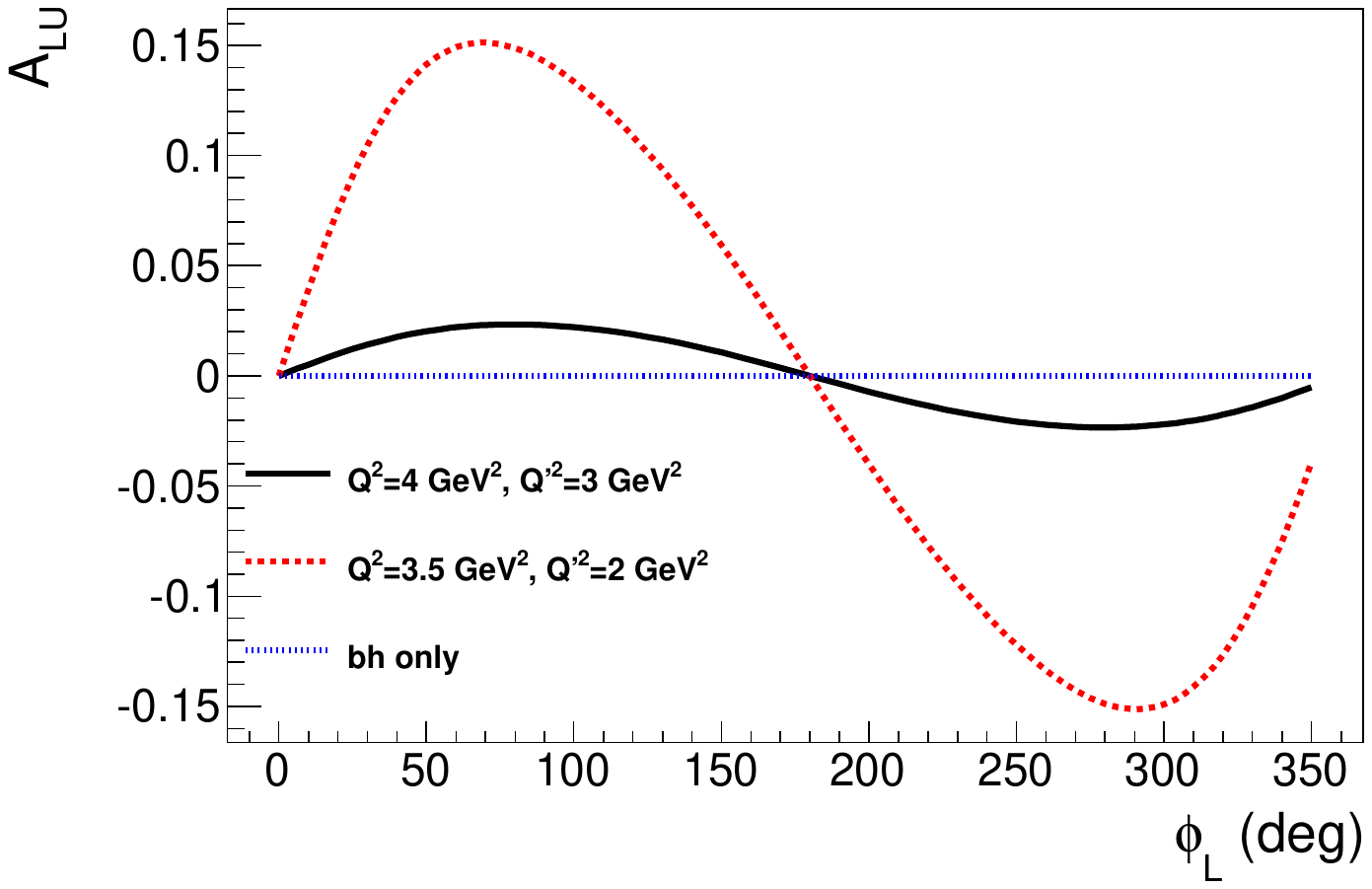}  \vspace{-0.5cm}
\caption{Out-of-plane angular dependence of the differential cross section (left) and the beam spin asymmetry (right) for the 
$eP\to e'p \mu^+\mu^-$ process at E=11~GeV, $x_B$=0.25, $t$=-0.4~GeV$^2$, and different virtual photon masses.}
\label{fig:ALU}
\end{center}
\end{figure}

\begin{figure}[!ht]
\begin{center}
\includegraphics [width=6cm]{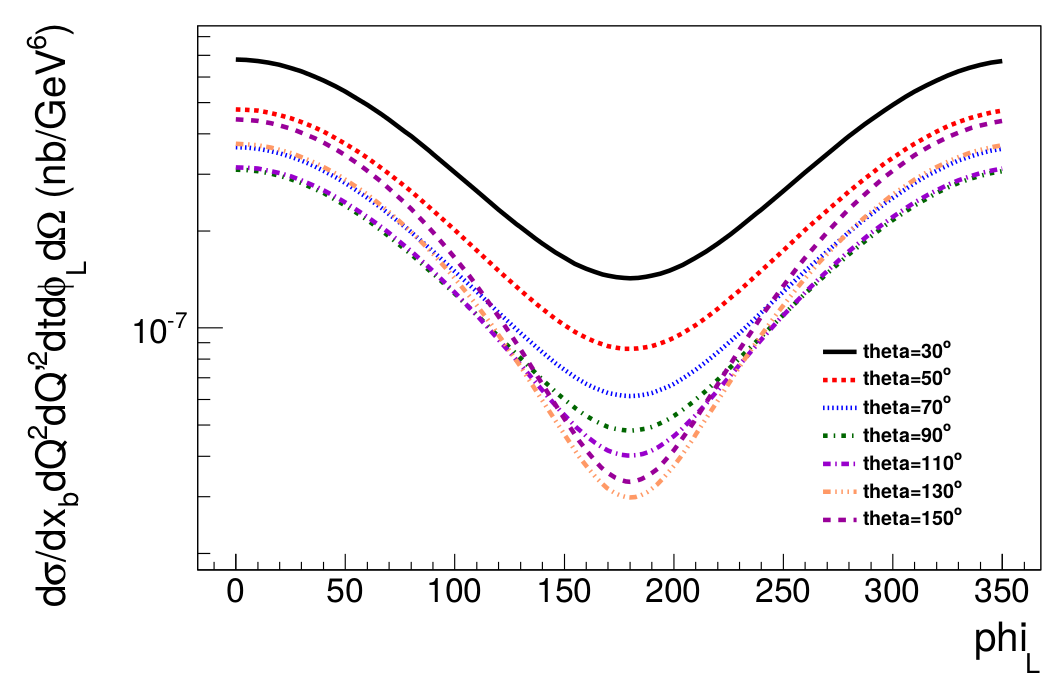} 
\includegraphics [width=6cm]{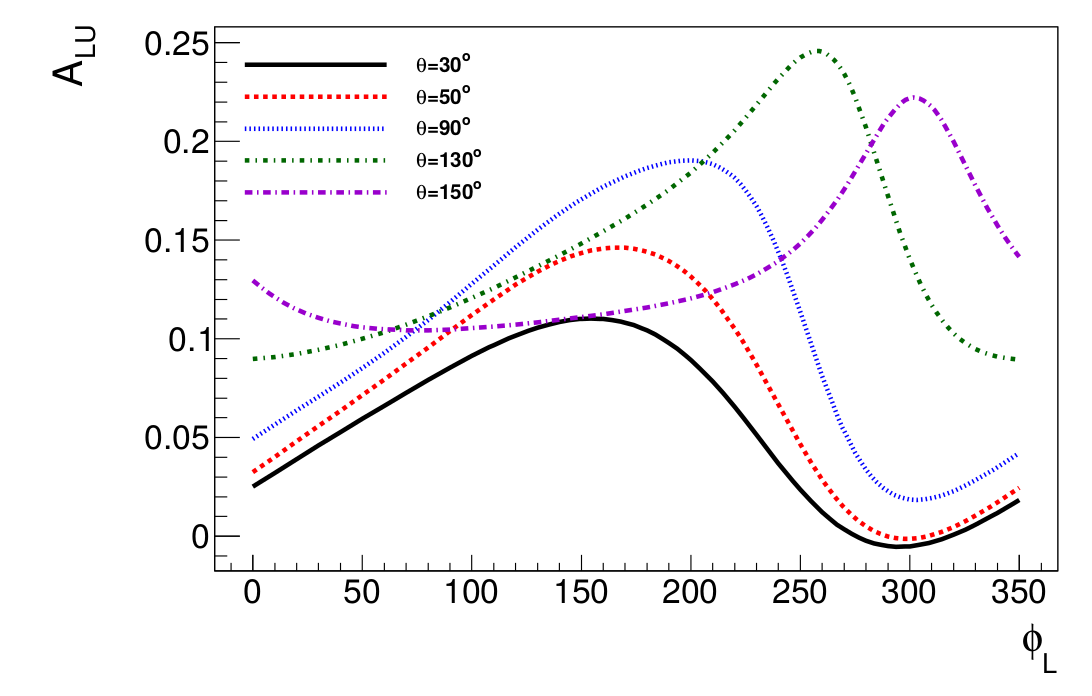}   \vspace{-0.2cm}
\caption{Out-of-plane angular dependence of the differential cross section for various polar angles of the muon (left) and for the differential beam spin asymmetry (right)}
\label{fig:p2}
\end{center}
\end{figure}

 We display in this section some of our projections made for JLab kinematics. Calculations are made from VGG model for   GPDs~\cite{Van99}, based on the DDVCS calculations from~\cite{Gui03}, with   modifications. Our kinematic studies are performed with our event generator. Fig.~\ref{fig:ALU} displays the differential cross section and the beam spin asymmetry $A_{LU}$ 
for two relevant kinematics for the determination of GPDs that we aim at measuring at JLab, in Hall A. These experimental observables are 
been obtained using the prescription of Eq.~\ref{eq:ALU}  
for the integration over the angular phase space of the di-muon pair. 
Sizeable asymmetries are predicted together with, as expected, a strong sensitivity of the cross section to kinematic conditions.  
Since the azimuthal and polar angle of the final muon pair are strongly correlated to each other and to the rate of BH (especially "BH2") in the full DDVCS+BH reaction,  we 
calculated the unpolarized cross section  (Fig.~\ref{fig:p2}, left) and beam spin asymmetry  (right)
at  different    $\theta_\mu$. 
The "actual" observables that will be used for the extraction of the CFFs are unpolarized cross sections and  beam spin asymmetries, differential in the initial and final azimuthal angles, integrated over the final polar angle. Extremes values in $\theta$ are cut out from this integral 
($40 \le \theta_\mu \le 140^0$), due to the large BH dominance in this regions coming from the peaks induced by the "BH2" diagrams. 

\section{Measuring DDVCS with the SoLID spectrometer at JLab Hall A}

\begin{figure}[ht!]
\centering
    \includegraphics[width=0.59\linewidth]{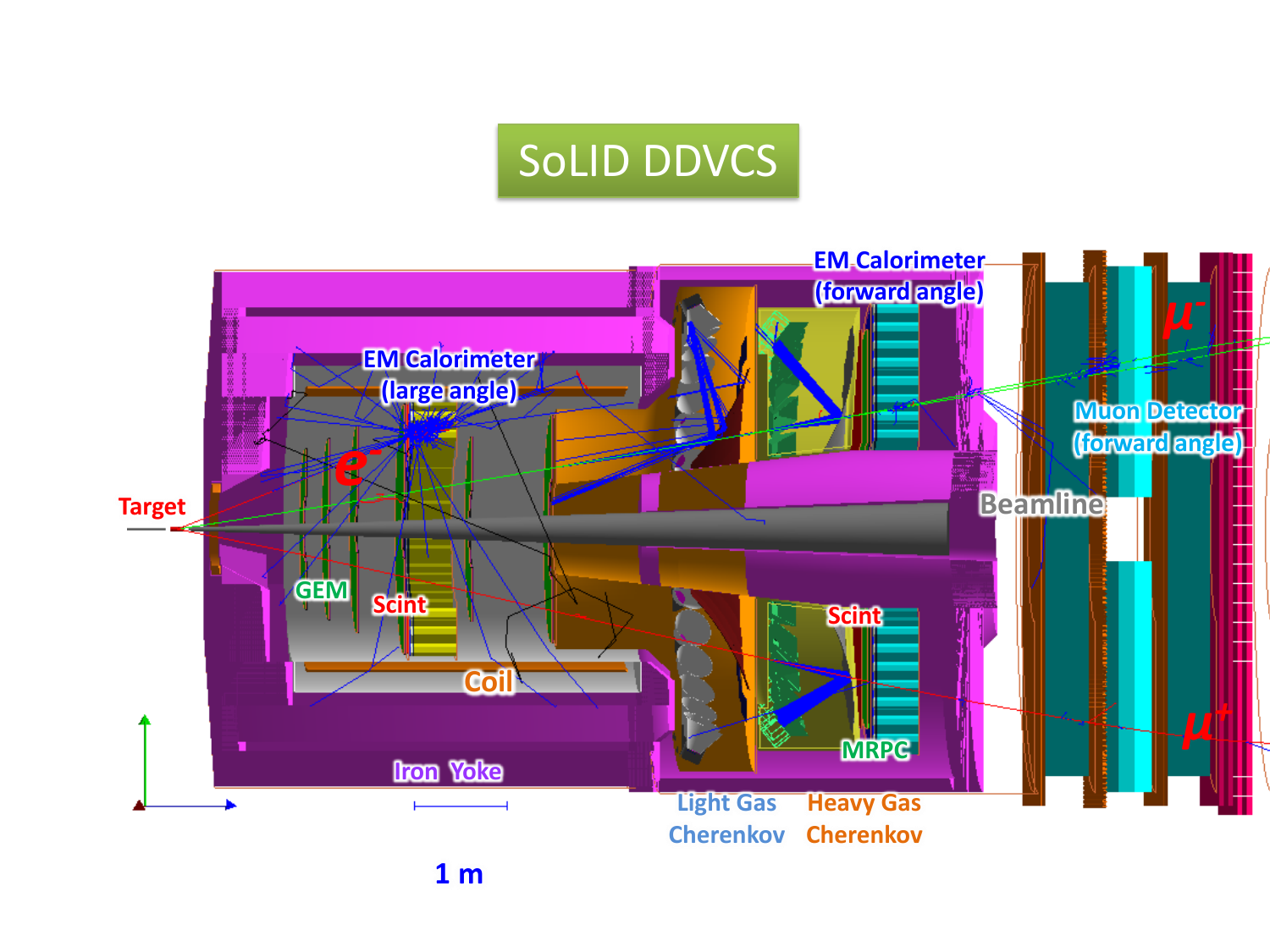}
    \includegraphics[width=0.35\textwidth]{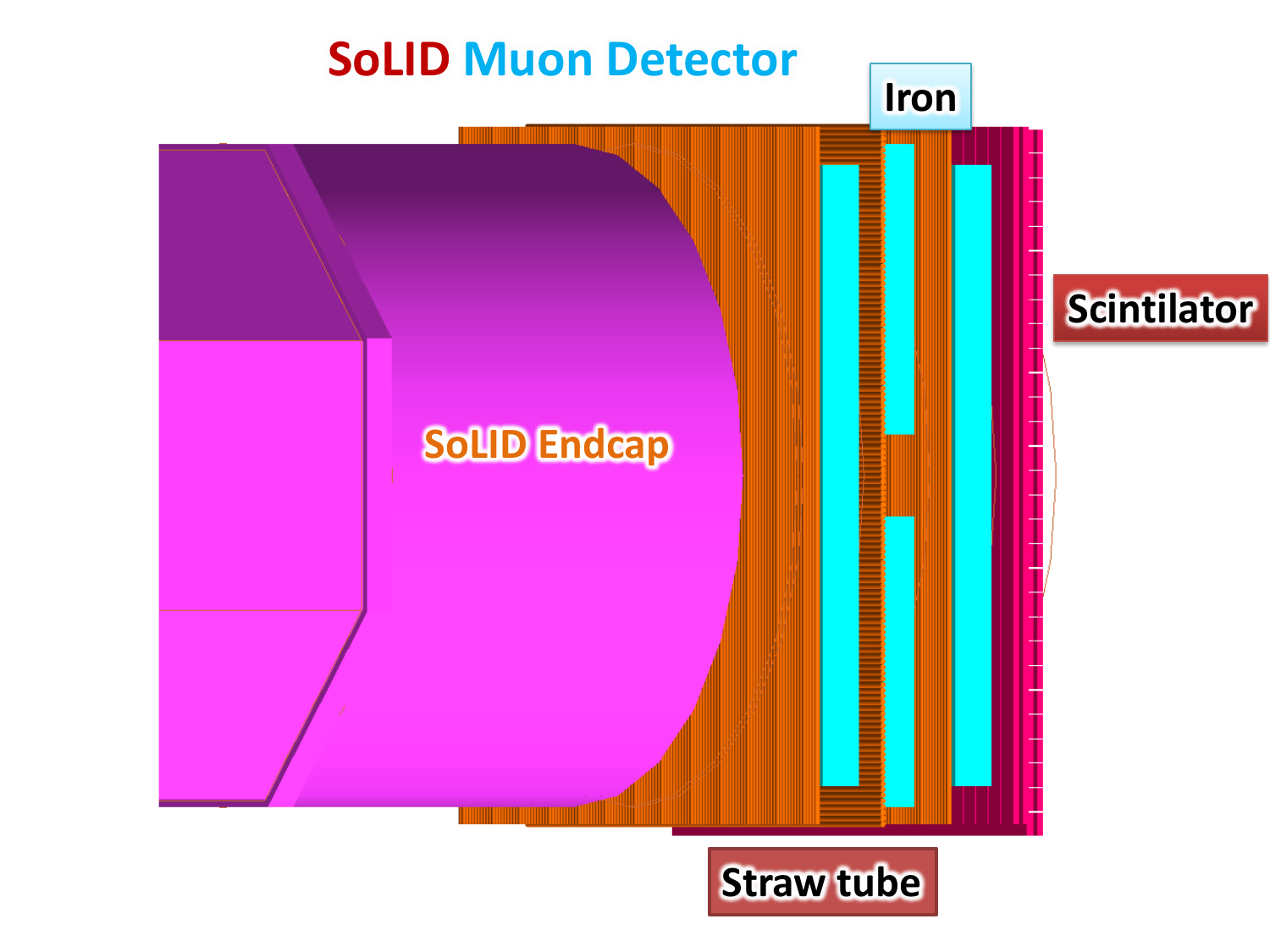}
\caption{\small{Left: SoLID DDVCS with the forward angle muon detector added to the $J\psi$ setup~\cite{jpsi}. Right: Preliminary design of SoLID muon detector at forward angle in Geant4 simulation.}}
\label{fig:solid}
\end{figure}

Different projects exist at JLab to measure DDVCS in Halls A, B, C. The project presented in this paper is based on our letter of intent~\cite{DDLOI2023}, submitted in 2023, and aiming at measuring the DDVCS+BH beam spin asymmetries with a slightly modified version of the SoLID (future) spectrometer. The projections presented in this section are based on this LOI. Our idea is to supplement the SoLID spectrometer~\cite{SoLID_pCDR}, as  in 
it's setup approved for the J/$\Psi$ experiment~\cite{jpsi}, with a dedicated muon detector, also serving as trigger. We are displaying our full setup, from GEANT4 simulations, Fig.~\ref{fig:solid} (left). 
The additional muon detector is shown Fig.~\ref{fig:solid} (right), and made of an alternance of 3 layers of iron (for shielding), 3 layers of straw tubes (for tracking), then 2 layers of plastic 
scintillators at the back-end to trigger on the muon pair. The iron plates  will be coming from CLEO-II. \smallskip

SoLID is designed to use a solenoid field and allows to carry out experiments 
 with high energy  $\sim11$ GeV electron beams on unpolarized and polarized targets 
at luminosities of up to $L=10^{37}$ cm$^{-2}$ sec$^{-1}$ in an open geometry. The solenoid field, reaching about 1.4T is provided 
by re-using the CLEO-II magnet.
Part of the CLEO-II iron flux return will be modified and reused, and two new
iron endcaps will be added at the front and back of the solenoid. 
Forward angle detectors cover 
polar angles from $8.5^\circ$ to $16^\circ$, and consist of planes of Gas
Electron Multipliers (GEM) for tracking, a light-gas Cherenkov (LGCC) for
e/$\pi$ separation, a heavy gas Cherenkov (HGCC) for $\pi$/K separation,
a Multi-gap Resistive Plate Chamber (MRPC) for time-of-flight, and an
Electromagnetic Calorimeter (FAEC). 
Large-angle detectors cover polar angle from $17^\circ$ to $24.5^\circ$ and consist of several planes of GEM for tracking, and an Electromagnetic Calorimeter (LAEC). 

\section{Summary}

We presented the motivations and our proposed setup for an experiment aiming at measuring DDVCS+BH with the SoLID spectrometer at JLab Hall A, supplemented by new muon detectors. 
The DDVCS+BH azymuthal dependencies of the beam spin asymmetries are sensitive to the imaginary part of the amplitudes and to the interference between DDVCS and BH. DDVCS has never been measured, 
 and accessing CFFs from its observables will enable, for the first time, to deconvolute the proton's and quark's momenta, which is essential for tomographic interpretations of the GPDs.
\medskip

Acknowledgements: The authors would like to thank, in alphabetic order, for their inputs or contributions, 
S. Alavarado, A. Camsonne, R. Gilman, M. Guidal, E. Voutier, S. Zhao, Z. Zhao. 
We also would like to thank all the members of the SoLID collaboration. 
This work is supported by the Virginia Tech Physics Department.

\end{document}